\documentclass[aps,prb,twocolumn,superscriptaddress,showpacs,amsmath,amssymb]{revtex4-1}

\usepackage{graphicx}

\begin{document}

\title{Quasi-two-dimensional Fermi surfaces with localized $f$ electrons in the layered heavy-fermion compound CePt$_2$In$_7$}

\author{K. G\"{o}tze}
\altaffiliation[Present address: ]{Department of Physics, University of Warwick, Gibbet Hill Road, Coventry, CV4 7AL, United Kingdom}
\affiliation{Hochfeld-Magnetlabor Dresden (HLD-EMFL), Helmholtz-Zentrum Dresden-Rossendorf and TU Dresden, D-01314 Dresden, Germany}

\author{Y. Krupko}
\affiliation{Laboratoire National des Champs Magn\'{e}tiques Intenses (LNCMI-EMFL), CNRS, UGA, 38042 Grenoble, France}

\author{J.A.N. Bruin}
\affiliation{High Field Magnet Laboratory (HFML-EMFL), Radboud University, 6525 ED Nijmegen, The Netherlands}

\author{J. Klotz}
\affiliation{Hochfeld-Magnetlabor Dresden (HLD-EMFL), Helmholtz-Zentrum Dresden-Rossendorf and TU Dresden, D-01314 Dresden, Germany}

\author{R.D.H. Hinlopen}
\affiliation{High Field Magnet Laboratory (HFML-EMFL), Radboud University, 6525 ED Nijmegen, The Netherlands}

\author{S. Ota}
\affiliation{Graduate School of Science and Technology, Niigata University, Niigata 950-2181, Japan}

\author{Y. Hirose}
\affiliation{Faculty of Science, Niigata University, Niigata 950-2181, Japan}

\author{H. Harima}
\affiliation{Graduate School of Science, Kobe University, Kobe 657-8501, Japan}

\author{R. Settai}
\affiliation{Faculty of Science, Niigata University, Niigata 950-2181, Japan}

\author{A. McCollam}
\affiliation{High Field Magnet Laboratory (HFML-EMFL), Radboud University, 6525 ED Nijmegen, The Netherlands}

\author{I. Sheikin}
\email[]{ilya.sheikin@lncmi.cnrs.fr}
\affiliation{Laboratoire National des Champs Magn\'{e}tiques Intenses (LNCMI-EMFL), CNRS, UGA, 38042 Grenoble, France}

\date{\today}

\begin{abstract}
We report measurements of the de Haas-van Alphen effect in the layered heavy-fermion compound CePt$_2$In$_7$ in high magnetic fields up to 35~T. Above an angle-dependent threshold field, we observed several de Haas-van Alphen frequencies originating from almost ideally two-dimensional Fermi surfaces. The frequencies are similar to those previously observed to develop only above a much higher field of 45~T, where a clear anomaly was detected and proposed to originate from a change in the electronic structure [M. M. Altarawneh \textit{et al}., Phys. Rev. B \textbf{83}, 081103 (2011)]. Our experimental results are compared with band structure calculations performed for both CePt$_2$In$_7$ and LaPt$_2$In$_7$, and the comparison suggests localized $f$ electrons in CePt$_2$In$_7$. This conclusion is further supported by comparing experimentally observed Fermi surfaces in CePt$_2$In$_7$ and PrPt$_2$In$_7$, which are found to be almost identical. The measured effective masses in CePt$_2$In$_7$ are only moderately enhanced above the bare electron mass $m_0$, from 2$m_0$ to 6$m_0$.
\end{abstract}

\maketitle

\section{Introduction}

Quantum critical points (QCPs) are a subject of considerable interest within the condensed-matter community. Heavy-fermion (HF) materials are particularly important in this context, as they can be conveniently tuned to QCPs by hydrostatic pressure, chemical doping, or magnetic field. Structural and magnetic dimensionality play a significant role in the physics of these materials, with reduced dimensionality understood to increase the strength of electronic correlations~\cite{Shishido2010,Mizukami2011}. Recently, a zero-temperature global phase diagram of heavy-fermion compounds with two types of quantum critical points was proposed, where the magnetic dimensionality is an important parameter~\cite{Si2006,Si2010a,Custers2012,Si2014}. Within this theoretical model, two-dimensionality favors the so-called local or Kondo-breakdown-type QCP~\cite{Coleman2001,Si2001,Si2003}, while a more conventional spin-density-wave-type QCP~\cite{Moriya1995,Hertz1976,Millis1993} is expected in three dimensions~\cite{}. One way to distinguish between these two types of QCPs is to determine whether the \(f\) electrons are itinerant or localized, i.e., whether or not they contribute to the Fermi surface (FS), on both sides of the QCP. This can be achieved by comparing experimentally established FS topology with the results of band structure calculations performed for both itinerant and localized \(f\) electrons. Magnetic quantum oscillations, such as the de Haas-van Alphen (dHvA) effect, are the most direct way to establish the FS topology of a metal. The reduced dimensionality of the FS is also expected to enhance unconventional superconductivity~\cite{Monthoux2003}, which is often observed in HF systems in the vicinity of a QCP. It has been demonstrated experimentally that the FS dimensionality is indeed one of the key parameters determining the superconducting critical temperature in HF materials~\cite{Goetze2015}, so precise knowledge of the topology, particularly the dimensionality, of the FS of HF materials close to a QCP is of primary importance.

Antiferromagnetic CeRhIn$_5$ with a tetragonal crystal structure and a N\'{e}el temperature $T_N =$ 3.8 K~\cite{Hegger2000} is one of the best-studied HF compounds. In this material, a QCP associated with the suppression of antiferromagnetism can be induced by either pressure~\cite{Knebel2006,Park2006} or high magnetic field~\cite{Jiao2015}. In the former case, dHvA frequencies change abruptly, and effective masses diverge exactly at the critical pressure $P_c \simeq$ 2.35 GPa required to suppress antiferromagnetic order~\cite{Shishido2005}. While the dHvA frequencies measured at ambient and low pressure correspond to localized $f$ electrons, those observed above $P_c$ suggest that the $f$ electrons contribute to the FS. Therefore, the pressure-induced QCP is of the Kondo-breakdown type. In high magnetic fields, on the other hand, new dHvA frequencies are reported to appear at $B^* =$ 30 T, well below the critical field $B_{c0} \approx$ 50 T to suppress antiferromagnetism~\cite{Jiao2015}. These new frequencies persist unchanged to fields above 50 T. Thus, the field-induced QCP in CeRhIn$_5$ appears to be of the spin-density-wave type.

CePt$_2$In$_7$ crystalizes into a body-centered tetragonal crystal structure~\cite{Kurenbaeva2008,Tobash2012,Klimczuk2014}. Remarkably, some details of the crystal structure of this compound, such as the exact position of the atoms and occupancy of the atomic sites, were initially wrongly assumed~\cite{Kurenbaeva2008,Tobash2012} and corrected only recently~\cite{Klimczuk2014}. Although the crystal structures of CePt$_2$In$_7$ and CeRhIn$_5$ are different, both compounds belong to the family of Ce$_nT_m$In$_{3n+2m}$ ($T$ is transition metal, $n =$ 1, 2, $m =$ 0, 1, 2), containing a sequence of $n$ CeIn$_3$ layers intercalated by $m$ $T$In$_2$ layers along the $c$ axis. In these systems, the FS dimensionality is expected to decrease with increasing distance between CeIn$_3$ layers. The building block of the series, CeIn$_3$ ($n = 1$, $m = 0$), crystallizes into a simple cubic structure and has an isotropic FS. The layered structures, with $m \neq$ 0, are characterized by strongly anisotropic properties and quasi-two-dimensional (quasi-2D) FS. Indeed, quasi-2D FS sheets were observed in the monolayer ($n = 1$, $m = 1$) system CeRhIn$_5$~\cite{Shishido2002,Hall2002}. In CePt$_2$In$_7$ ($n = 1$, $m = 2$), where CeIn$_3$ layers are separated by two PtIn$_2$ layers, the FS is expected to be more 2D.

The physical properties of CePt$_2$In$_7$ and CeRhIn$_5$ also have a lot of similarities. CePt$_2$In$_7$ orders antiferromagnetically at a N\'{e}el temperature $T_N =$ 5.5 K~\cite{Bauer2010a}, slightly higher than that of CeRhIn$_5$. Similar to that of CeRhIn$_5$, the antiferromagnetic order in CePt$_2$In$_7$ is suppressed by pressure, and a pressure-induced QCP was reported to occur at a critical pressure $P_c =$ 3.2--3.5 GPa~\cite{Bauer2010a,Sidorov2013,Kurahashi2015}. Superconductivity with a critical temperature $T_c =$ 2.1 K at $P_c$~\cite{Sidorov2013} was observed in both polycrystalline samples~\cite{Bauer2010a} and single crystals~\cite{Sidorov2013,Kurahashi2015}. The critical temperature is similar to that observed in CeRhIn$_5$ and is one of the highest among Ce-based HF materials. It was recently reported that $T_N$ in CePt$_2$In$_7$ is also suppressed by a magnetic field applied along either the $a$ or $c$ axis, possibly giving rise to a QCP slightly below 60 T~\cite{Krupko2016}, again in close similarity to CeRhIn$_5$~\cite{Jiao2015}.

Previously reported magnetic quantum oscillation measurements in CePt$_2$In$_7$ performed in pulsed magnetic fields by the tunnel-diode-oscillator technique~\cite{Altarawneh2011} suggest that the compound comes closer to realizing a 2D analog of CeIn$_3$ than CeRhIn$_5$. In these measurements, only small and almost isotropic FS pockets, with dHvA frequencies below 2~kT, exhibiting field-dependent effective masses were observed below a distinct anomaly that occurs at $B_m =$ 45 T. Above $B_m$, however, much higher dHvA frequencies corresponding to three almost ideally 2D FS sheets were detected. A feature similar to $B_m$ was previously observed in CeIn$_3$ at about the same field~\cite{Purcell2009}, where the dHvA frequencies also change. More recently, however, the same high dHvA frequencies were observed below $B_m$ in CePt$_2$In$_7$~\cite{Miyake2015}. This questions a change in the dHvA frequencies at $B_m$ in CePt$_2$In$_7$, and, therefore, the degree of similarity between this compound and CeIn$_3$. These conflicting results demand further investigation. Furthermore, it is still not clear whether a Fermi surface reconstruction occurs in CePt$_2$In$_7$ deep inside the antiferromagnetic phase, as in CeRhIn$_5$~\cite{Jiao2015}, or the observation of the high dHvA frequencies below $B_m$ is due to magnetic breakdown.

Another open question is whether the $f$ electrons are itinerant or localized in CePt$_2$In$_7$, especially at zero and low magnetic fields. The angle dependence of the dHvA frequencies previously observed above $B_m$ were found to be in good agreement with the results of band structure calculations in which the $f$ electrons are confined mostly to their atomic cores~\cite{Altarawneh2011}. The observation of the same frequencies below $B_m$~\cite{Miyake2015} suggests that the $f$ electrons are also localized at low magnetic fields. However, these band structure calculations were performed assuming the wrong crystal structure of CePt$_2$In$_7$. More recently, it was demonstrated that the band structure calculations assuming the correct crystal structure result in different topology of the FS~\cite{Klimczuk2014}. Finally, angle-resolved photoemission spectroscopy measurements on CePt$_2$In$_7$ were very recently reported~\cite{Shen2017}, and the results were compared with band structure calculations assuming itinerant $f$ electrons. Most of the observed electronic features could be accounted for by the calculations, although some discrepancies were also found.

In this paper, we report dHvA measurements in CePt$_2$In$_7$ and PrPt$_2$In$_7$, which is known to have localized $f$ electrons, in magnetic fields up to 35 T. The dominant observed FSs in both compounds are almost ideally two-dimensional, as is expected from their crystal structure. The FSs of the two compounds are almost identical and are well described by the band structure calculations performed for LaPt$_2$In$_7$, implying localized $f$ electrons in CePt$_2$In$_7$. The effective masses we observe in CePt$_2$In$_7$, in fields up to 35 T, are only moderately enhanced above a bare electron mass but are, nevertheless, somewhat larger than those observed at much higher magnetic fields above $B_m$ in Ref.~\onlinecite{Altarawneh2011}.

\begin{figure}[htb]
\includegraphics[width=7.5cm]{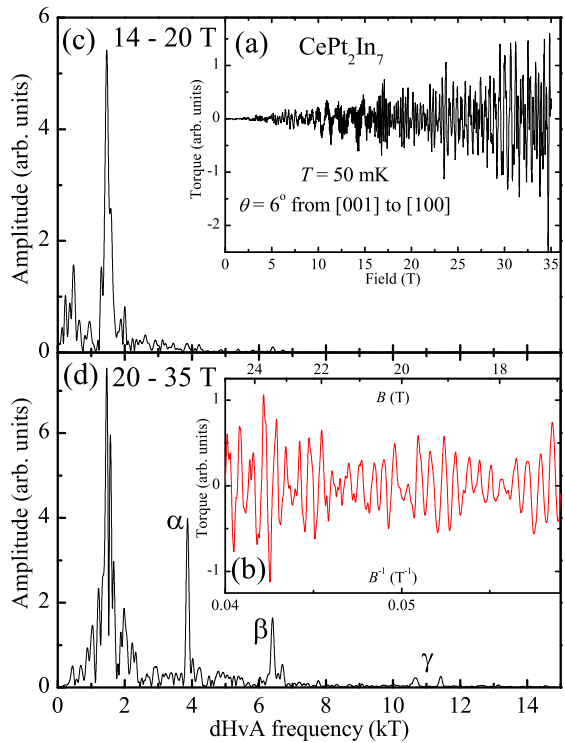}
\caption{\label{CePt2In7_dHvA} (a) The oscillatory torque signal after subtracting the nonoscillating background in CePt$_2$In$_7$ for magnetic field applied at 6$^\circ$ off the $c$ axis at 50 mK. (b) A high-field zoom of the dHvA oscillations from (a) plotted as a function of the inverse magnetic field. (c) and (d) FFT spectra of the dHvA oscillations from (a) over equal $1/B$ intervals below (14--20 T) and above (20--35 T) 20~T, respectively.}
\end{figure}

\begin{figure}[htb]
\includegraphics[width=7.5cm]{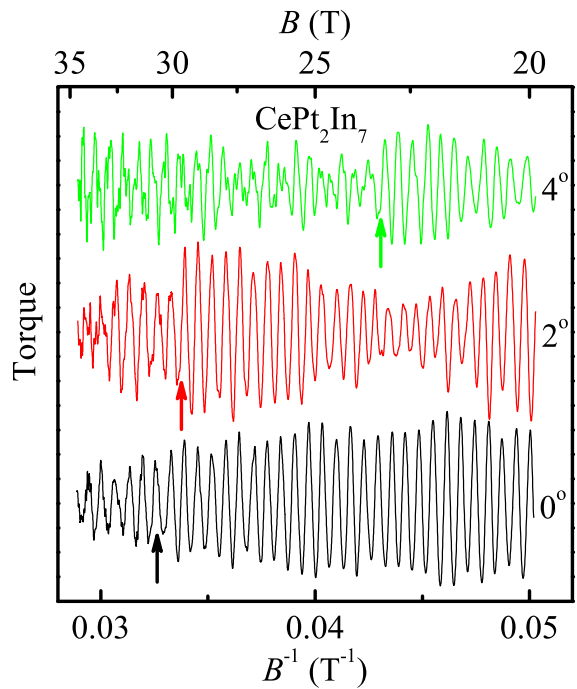}
\caption{\label{Small_angles} dHvA oscillations in CePt$_2$In$_7$ for magnetic field applied at several small angles from the $c$ towards the $a$ axis at 50 mK. The arrows indicate approximately the field above which high dHvA frequencies start to develop for each orientation of the magnetic field.}
\end{figure}

\begin{figure*}[htb]
\includegraphics[width=15cm]{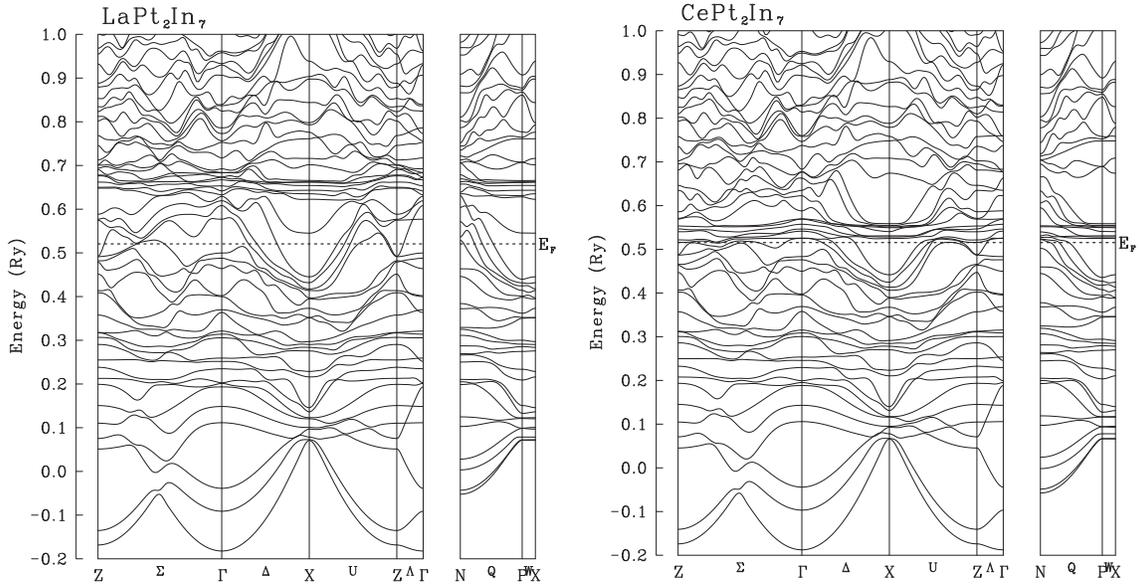}
\caption{\label{band_structure}Calculated band structure along the symmetry axes for LaPt$_2$In$_7$ (left) and CePt$_2$In$_7$ (right). The Fermi level is denoted by $E_F$.}
\end{figure*}

\section{Experimental details}

High quality single crystals of CePt$_2$In$_7$ and PrPt$_2$In$_7$ used for the dHvA measurements reported here were grown by an In self-flux method, and the details are given elsewhere~\cite{Kurahashi2015}. The resulting single crystals are small thin platelets with the crystallographic $c$ axis perpendicular to their large surfaces. The CePt$_2$In$_7$ and PrPt$_2$In$_7$ samples were prepared using the same procedure. Measurements of the dHvA effect were performed using a conventional torque magnetometry technique. The measurements on CePt$_2$In$_7$ were carried out using either a metallic cantilever in a top-loading dilution refrigerator in fields up to 35~T or a piezoresistive microcantilever in a $^3$He cryostat up to 33~T. The CePt$_2$In$_7$ sample mounted on the metallic cantilever is the same single crystal used for previous specific-heat measurements~\cite{Krupko2016}, which confirmed the absence of any impurity phases. A smaller sample from the same batch was chosen for microcantilever measurements. The dHvA measurements on a PrPt$_2$In$_7$ sample mounted on a metallic cantilever were performed either in a top-loading dilution refrigerator in fields up to 35~T or in a variable-temperature insert in a superconducting magnet up to 16~T.

\begin{figure}[htb]
\includegraphics[width=7.5cm]{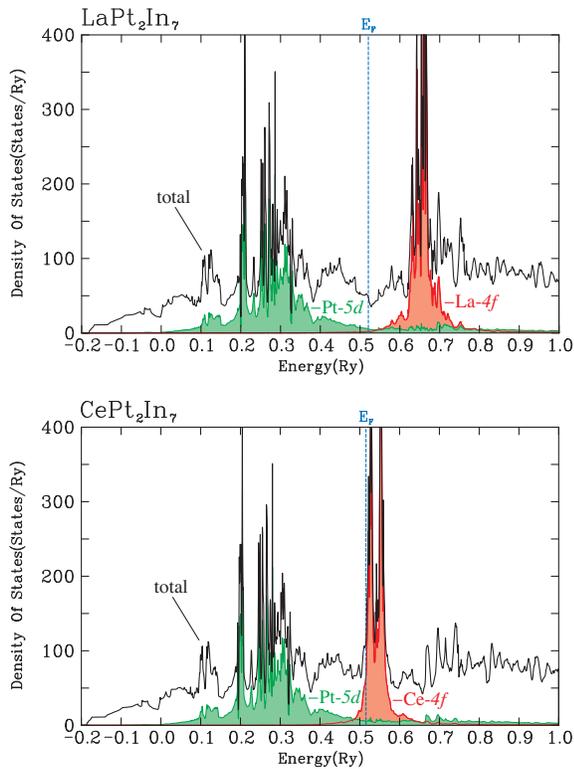}
\caption{\label{dos}Calculated density of states for LaPt$_2$In$_7$ (top) and CePt$_2$In$_7$ (bottom). $E_F$ denotes the Fermi level.}
\end{figure}

\begin{figure*}[htb]
\includegraphics[width=15cm]{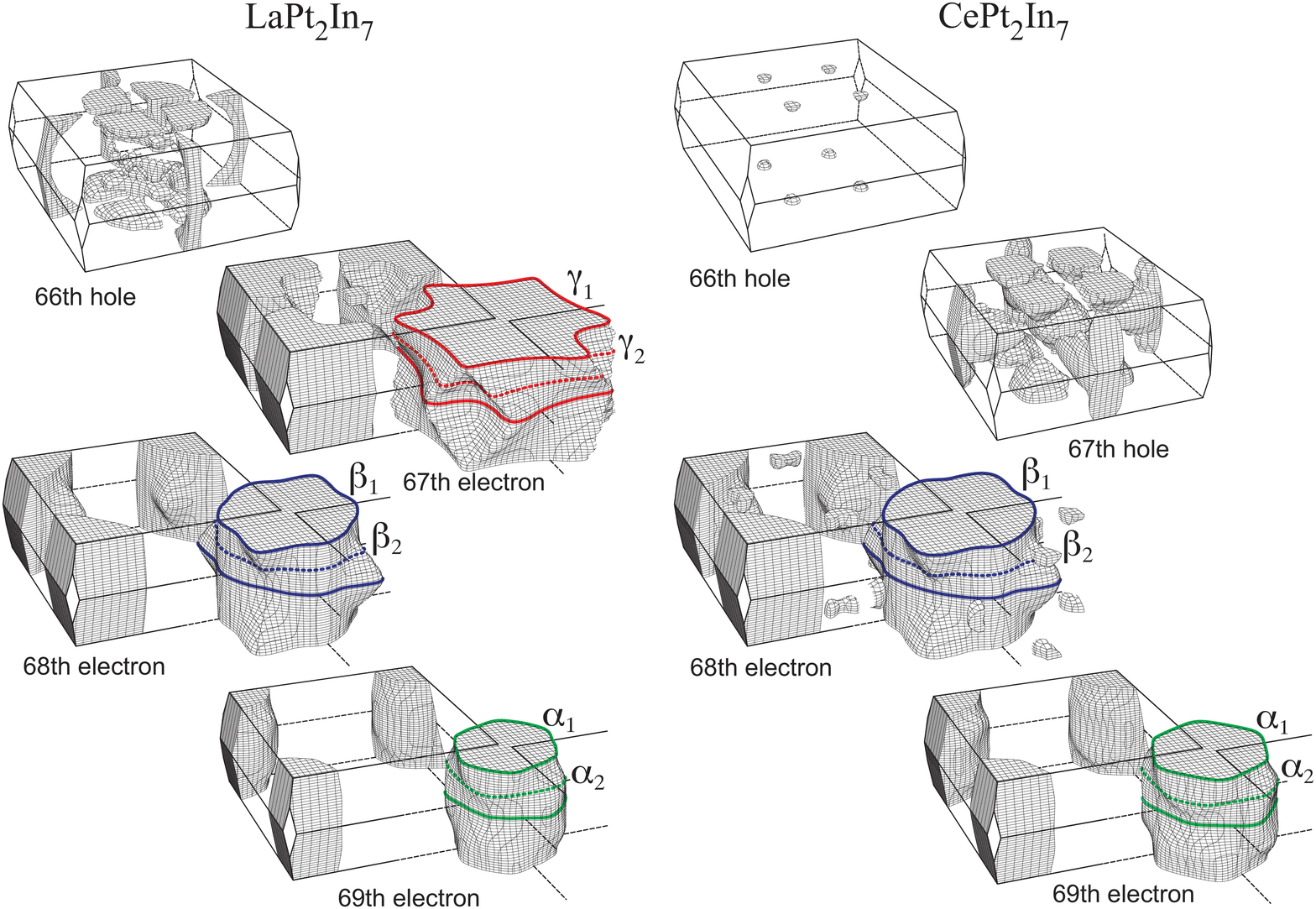}
\caption{\label{FS}Calculated FSs of LaPt$_2$In$_7$ (left) and CePt$_2$In$_7$ (right). Solid and dotted lines indicate the two extreme cross sections of the main FS sheets of LaPt$_2$In$_7$. These orbits give rise to the experimentally observed dHvA frequencies when the magnetic field is applied along the $c$ axis. }
\end{figure*}

\section{Results and discussion}

\subsection{dHvA oscillations}

Figure~\ref{CePt2In7_dHvA}(a) shows the oscillatory torque after subtracting a nonoscillating background in CePt$_2$In$_7$ over the whole field range up to 35~T. The oscillations appear at fields as low as 2~T, confirming the high quality of our sample. However, only relatively low dHvA frequencies, up to 2~kT, are observed in magnetic fields up to about 20~T. Above this field, additional higher dHvA frequencies start to develop, as can be seen in Fig.~\ref{CePt2In7_dHvA}(b), which shows a high-field zoom of the oscillations. This is further confirmed by the fast Fourier transform (FFT) spectra of the oscillations below and above 20~T shown in Figs.~\ref{CePt2In7_dHvA}(c) and~\ref{CePt2In7_dHvA}(d), respectively. Three high fundamental frequencies, denoted $\alpha$, $\beta$, and $\gamma$, are observed when $B$ is applied at 6$^\circ$ from the $c$ to the $a$ axis above 20~T. The values of the high dHvA frequencies we observe here are in good agreement with those reported by Altarawneh \textit{et al}~\cite{Altarawneh2011}. In their work, the high frequencies were observed only above $B_m \approx$ 45~T, where a distinct anomaly in the tunnel-diode-oscillator signal occurs~\cite{Altarawneh2011}. In our measurements the high dHvA frequencies emerge at a much lower field.

It would be tempting to ascribe the appearance of the high dHvA frequencies to a FS modification associated with a change in the $f$ electron character from localized to itinerant, similar to the case for CeRhIn$_5$~\cite{Jiao2015}. However, in CeRhIn$_5$ the higher frequencies appear suddenly above $B^* \approx 30$ T, which does not depend on the orientation of the magnetic field. In CePt$_2$In$_7$, on the contrary, the high-frequency oscillations develop above a certain value of the magnetic field, which is strongly field orientation dependent, decreasing as the field is tilted away from the $c$ axis, as shown in Fig.~\ref{Small_angles}. Furthermore, as will be shown in the following, the angle dependence of the observed high dHvA frequencies unambiguously suggests that they originate from the FSs with localized $f$ electrons.

The most likely scenario is that the high frequencies appear at a relatively high field due to the development of magnetic breakdown orbits that span multiple FSs. Indeed, the antiferromagnetic ordering modifies the original crystallographic Brillouin zone. The magnetic structure of CePt$_2$In$_7$~\cite{Raba2017,Gauthier2017} implies that its antiferromagnetic Brillouin zone is 8 times smaller than the paramagnetic Brillouin zone. This results in fragmentation of the large FSs that exceed the size of the antiferromagnetic Brillouin zone, $\simeq 5320$~T. Most of the low dHvA frequencies observed at low magnetic fields are likely to originate from such fragmented FSs. The larger orbits can, however, be recovered through a magnetic breakdown. For such an anisotropic system as CePt$_2$In$_7$ the magnetic breakdown field is also expected to be strongly anisotropic, which is in keeping with our results. This scenario is further supported by recent specific-heat measurements suggesting that an anisotropic spin-density wave opens a gap on almost the entire FS below the N\'{e}el temperature at zero magnetic field~\cite{Krupko2016}. Indeed, gapped FSs can be observed in quantum oscillation measurements only at magnetic fields high enough to ensure the tunneling of quasiparticles through the gap, i.e., the magnetic breakdown.

\begin{figure*}[htb]
\includegraphics[width=15cm]{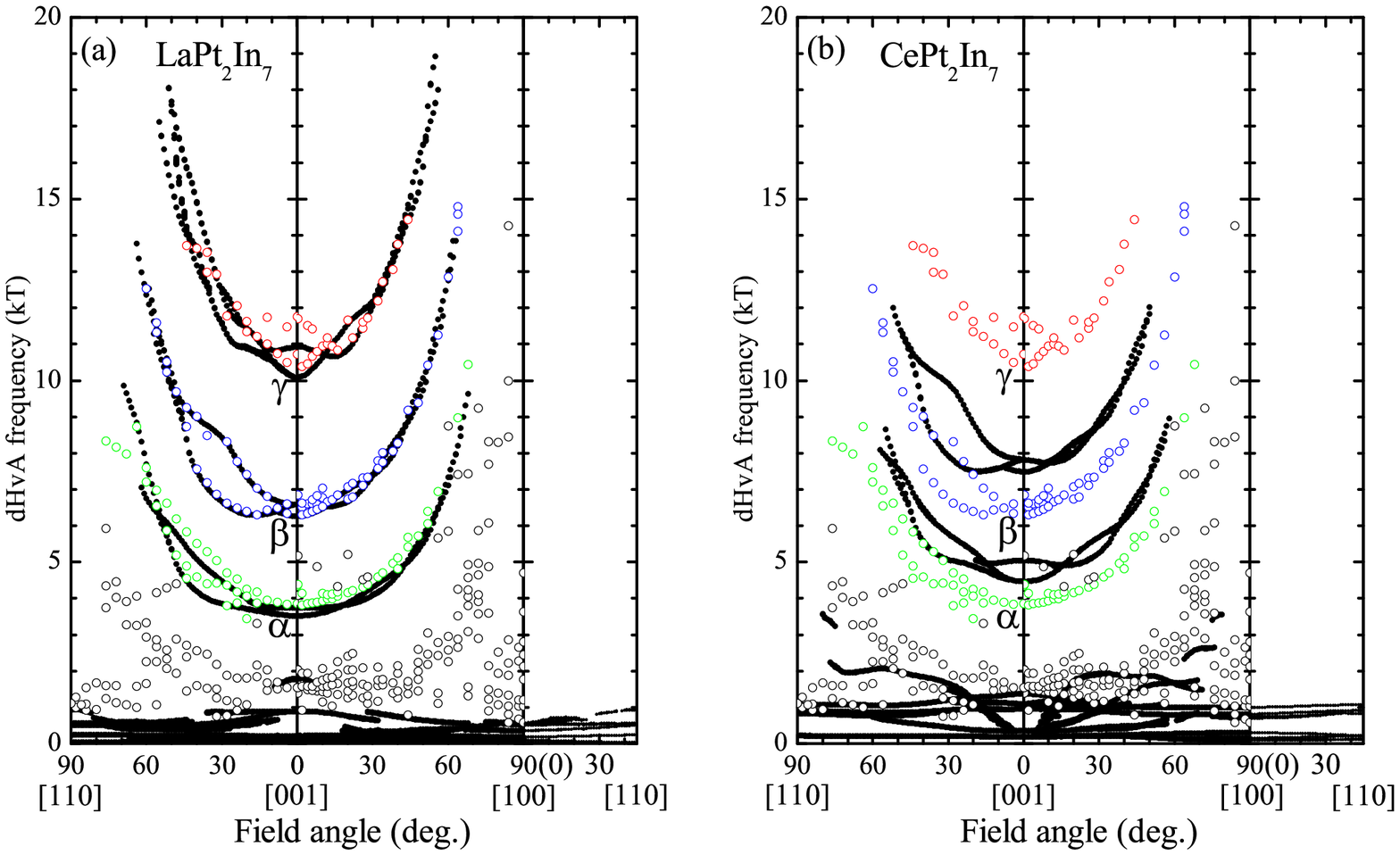}
\caption{\label{AngDep} The angle dependence of the experimentally observed dHvA frequencies in CePt$_2$In$_7$ (open circles) is shown together with the results of band structure calculations (solid circles) performed for (a) LaPt$_2$In$_7$ and (b) CePt$_2$In$_7$. The former corresponds to CePt$_2$In$_7$ with localized $f$ electrons.}
\end{figure*}

\subsection{Band structure calculations}

To ascertain whether the $f$ electrons are itinerant or localized in CePt$_2$In$_7$, we performed band structure calculations for both CePt$_2$In$_7$ and LaPt$_2$In$_7$, with the latter corresponding to localized $f$ electrons. The electronic band structure of both compounds was calculated within the local-density approximation (LDA) by using a full-potential linearized augmented plane-wave method. For the LDA, the formula proposed by Gunnarsson and Lundqvist~\cite{Gunnarsson1976} was used. The calculations were performed using the program codes TSPACE and KANSAI-13. Scalar relativistic effects are taken into account for all electrons, and spin-orbit interactions are included self-consistently for all valence electrons, as in a second variational procedure.

The space group of CePt$_2$In$_7$ is $I4/mmm$ (number 139 D$_{4h}^{17}$)~\cite{Klimczuk2014}. The lattice parameters used for the calculation are $a=4.6147\:${\AA} and $c=21.6510\:${\AA}, $z=0.32561$ for 4$e$ positions of Pt atoms, and $z=0.10781$ for 8$g$ positions of In atoms~\cite{Haga}. These parameters are similar to those reported by Klimczuk \textit{et al}.~\cite{Klimczuk2014}. There are no position parameters for Ce atoms in the 2$b$ position, or for In atoms in 2$a$ and 4$d$ positions. As single crystals of LaPt$_2$In$_7$ are presently unavailable, the same lattice parameters were used for LaPt$_2$In$_7$ calculations. Muffin-tin (MT) radii are set as 0.3542$a$ for Ce(La), and 0.2806$a$ for both Pt and In. Core electrons [Xe core minus 5$s^2$5$p^6$ for Ce(La), Xe core minus 5$p^6$ plus 4$f^{14}$ for Pt, and Kr core plus 4$d^{10}$ for In] are calculated inside the MT spheres in each self-consistent step. The 5$s^2$5$p^6$ electrons of Ce(La), 5$p^6$ electrons of Pt, and 4$d^{10}$ electrons of In are calculated as valence electrons by using the second energy window.

The linearized augmented plane-wave (LAPW) basis functions were truncated at  \(|\mbox{\boldmath$k + G$}_i| < 4.85 \times 2\pi /a\), corresponding to 1115 LAPW functions at the $\Gamma$ point. The 308 sampling $k$-points, which are uniformly distributed in the irreducible 16th of the Brillouin zone (divided by 20, 20, and 4), are used both for the potential convergence and for the final band structure. The calculated band structure and the density of states are shown in Figs.~\ref{band_structure} and~\ref{dos}, respectively. The calculated FSs are shown in Fig.~\ref{FS}.

\subsection{Comparison with band structure calculations}

As LaPt$_2$In$_7$ and CePt$_2$In$_7$ have the same layered crystal structure, it is expected that some of the calculated FS sheets are quasi-2D in both cases. Indeed, the quasi-2D FSs originating from bands 68 and 69, giving rise to orbits $\alpha$ and $\beta$, respectively, are very similar in both compounds, the main difference being that they are slightly smaller in the La compound. The details of the other FS sheets are, however, different. While band 66 contains only one tiny pocket for CePt$_2$In$_7$, there are several large pockets in the case of LaPt$_2$In$_7$. The most essential difference, however, is the topology of the FSs originating from the hole band 67. For CePt$_2$In$_7$, this band contains two types of moderate-size pockets that give rise to relatively small orbits. On the contrary, for LaPt$_2$In$_7$, the quasi-2D FS of band 67 gives rise to very large orbits $\gamma_1$ and $\gamma_2$. This difference alone allows us to decide which of the calculated FSs yields a better agreement with the experimental results.

Figure~\ref{AngDep}(a) shows the angle dependence of the dHvA frequencies observed in CePt$_2$In$_7$ together with the results of the band structure calculations for LaPt$_2$In$_7$, which correspond to CePt$_2$In$_7$ with localized $f$ electrons. Experimental and calculated frequencies and effective masses are also shown in Table~\ref{dHvA parameters}. The agreement between the experimentally observed $\alpha$, $\beta$, and $\gamma$ branches and those of the calculations is excellent. Not only is the shape of the curves the same, but the absolute values are almost identical. This implies that both the topology and the size of the calculated FS sheets reproduce well the experimental results. There are a number of lower dHvA frequencies observed in the experiment which are not predicted by the band structure calculations. These low frequencies are probably due to the modification of the Brillouin zone in the antiferromagnetic state, as all the calculations were performed for a paramagnetic ground state. A similar situation with unpredicted low frequencies was also observed in CeRhIn$_5$~\cite{Shishido2002}.

Contrary to the 4$f$-localized case, the agreement between the calculations for CePt$_2$In$_7$ with itinerant $f$ electrons and the experimental results is not nearly as good, as can be seen in Fig.~\ref{AngDep}(b). The experimentally observed $\alpha$ and $\beta$ branches are shifted downwards compared to the calculated ones (see also Table~\ref{dHvA parameters}). Most important, however, is the absence of the $\gamma$ branch in the band structure calculations with itinerant $f$ electrons.

The comparison of the experimentally observed angle dependence of the dHvA frequencies with the results of the band structure calculations strongly suggests that the $f$ electrons are localized in CePt$_2$In$_7$.

\begin{figure}[htb]
\includegraphics[width=7.5cm]{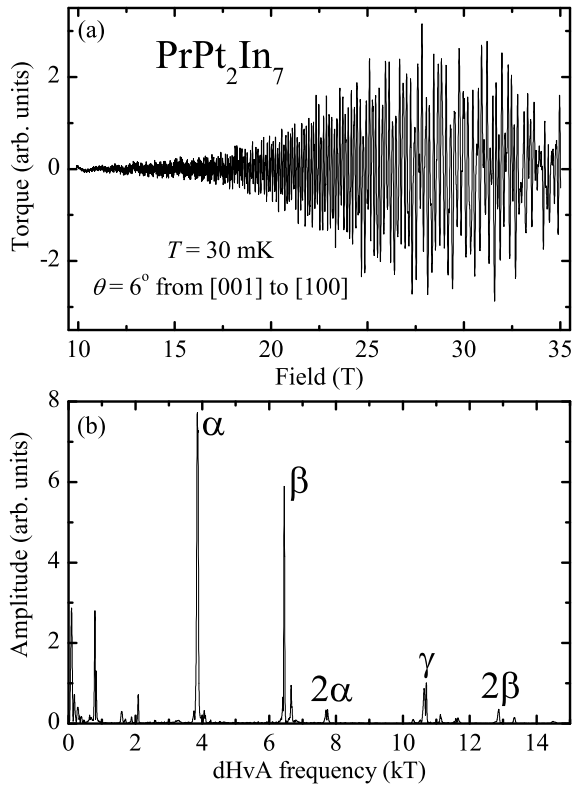}
\caption{\label{PrPt2In7_dHvA} High-field dHvA oscillations (a) and their Fourier spectrum (b) in PrPt$_2$In$_7$ for magnetic field applied at 6$^\circ$ from the $c$ axis at 30 mK.}
\end{figure}

\begin{figure}[htb]
\includegraphics[width=\columnwidth]{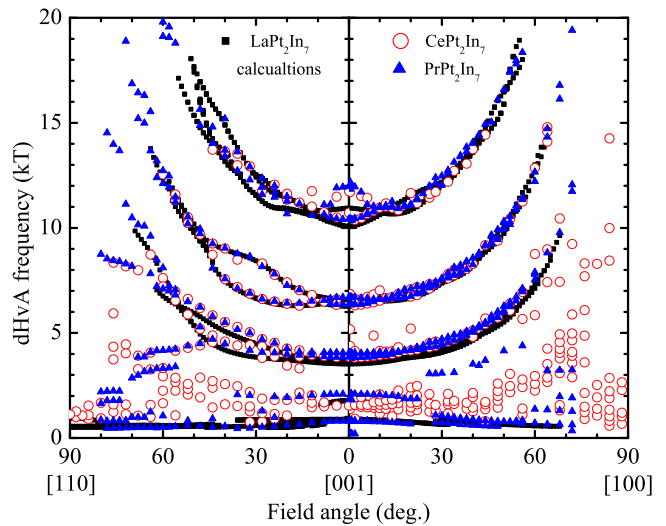}
\caption{\label{Pr_vs_Ce} Comparison of the experimentally observed angle dependence of the dHvA oscillations in CePt$_2$In$_7$ (open circles) and PrPt$_2$In$_7$ (solid triangles) with the result of the calculations for LaPt$_2$In$_7$ (solid squares). Calculated dHvA frequencies lower than 500~T are not shown for clarity.}
\end{figure}

\begin{table*}[htb]
\caption{\label{dHvA parameters}Experimentally determined dHvA frequencies and effective masses of CePt$_2$In$_7$ and PrPt$_2$In$_7$ for the magnetic field applied at 6$^\circ$ from the [001] to [100] direction. Calculated dHvA frequencies and band masses for both LaPt$_2$In$_7$ (localized $f$ electrons) and CePt$_2$In$_7$ (itinerant $f$ electrons) are also shown for comparison. Branch assignments refer to Fig.~\ref{AngDep}.}
\begin{ruledtabular}
\begin{tabular}{lcccccccc}
&\multicolumn{4}{c}{Experiment}&\multicolumn{4}{c}{Calculations}\\
\cline{2-5} \cline{6-9}
&\multicolumn{2}{c}{CePt$_2$In$_7$}&\multicolumn{2}{c}{PrPt$_2$In$_7$}&\multicolumn{2}{c}{LaPt$_2$In$_7$}&\multicolumn{2}{c}{CePt$_2$In$_7$}\\
\cline{2-3} \cline{4-5} \cline{6-7} \cline{8-9}
Branch& $F$ (kT)& $m^{\ast}/m_0$& $F$ (kT)& $m^{\ast}/m_0$& $F$ (kT)& $m_b/m_0$& $F$ (kT)& $m_b/m_0$\\
\hline
$\alpha_1$& 3.87& 2.27$\pm$0.04& 3.85& 0.69$\pm$0.02& 3.55& 0.49& 4.52& 1.29\\
$\alpha_2$& 4.03& & 4.06& & 3.75& 0.55& 5.01& 1.58\\
$\beta_1$& 6.40& 5.35$\pm$0.06& 6.41& 1.1$\pm$0.1& 6.35& 0.62& 7.58& 1.47\\
$\beta_2$& 6.60& & 6.66& & 6.55& 0.65& 7.76& 1.46\\
$\gamma_1$& 10.67& 5.1$\pm$0.2& 10.64& 1.4$\pm$0.1& 10.32& 1.14& &\\
$\gamma_2$& 11.43& 6.2$\pm$0.3& 11.12& 1.5$\pm$0.2& 10.84& 1.06& &\\
\end{tabular}
\end{ruledtabular}
\end{table*}

\subsection{Comparison with PrPt$_2$In$_7$}

The most direct way to confirm that the $f$ electrons are localized in CePt$_2$In$_7$ would be to compare the dHvA frequencies and their angle dependence in both CePt$_2$In$_7$ and LaPt$_2$In$_7$. This is, however, not possible since LaPt$_2$In$_7$ single crystals are currently unavailable, as was already mentioned above. We therefore performed dHvA measurements in PrPt$_2$In$_7$. As the $f$ electrons of Pr are known to be well localized, the FS of PrPt$_2$In$_7$ is expected to be nearly the same as that of LaPt$_2$In$_7$.

Figure~\ref{PrPt2In7_dHvA} shows the oscillatory torque, after subtracting a nonoscillating background, and the corresponding Fourier transform in PrPt$_2$In$_7$ for magnetic field applied at 6$^\circ$ from $c$ to the $a$ axis, the same orientation as for the data in Fig.~\ref{CePt2In7_dHvA}. The same fundamental frequencies, $\alpha$, $\beta$, and $\gamma$, are observed at the same values as in CePt$_2$In$_7$. The main difference in the FFT spectra of the two compounds is the number of low frequencies, which is much lower in PrPt$_2$In$_7$. This implies that the main large FSs of the two compounds are almost identical, but there are many more small pockets in CePt$_2$In$_7$ compared to PrPt$_2$In$_7$.

In Fig.~\ref{Pr_vs_Ce}, we compare the angle dependence of the experimentally observed dHvA frequencies in both CePt$_2$In$_7$ and PrPt$_2$In$_7$ with the calculated angle dependence for LaPt$_2$In$_7$, which corresponds to the case with localized $f$ electrons. The excellent agreement of the experimentally measured angle dependence of the main dHvA frequencies $\alpha$, $\beta$, and $\gamma$ observed in CePt$_2$In$_7$ and PrPt$_2$In$_7$ gives further confirmation that the $f$ electrons are localized in CePt$_2$In$_7$. Moreover, the comparison suggests that most of the low dHvA frequencies observed in CePt$_2$In$_7$ correspond to small FS pockets originating from the fragmentation of the FS upon the formation of the antiferromagnetic Brillouin zone.

\begin{figure}[htb]
\includegraphics[width=7.5cm]{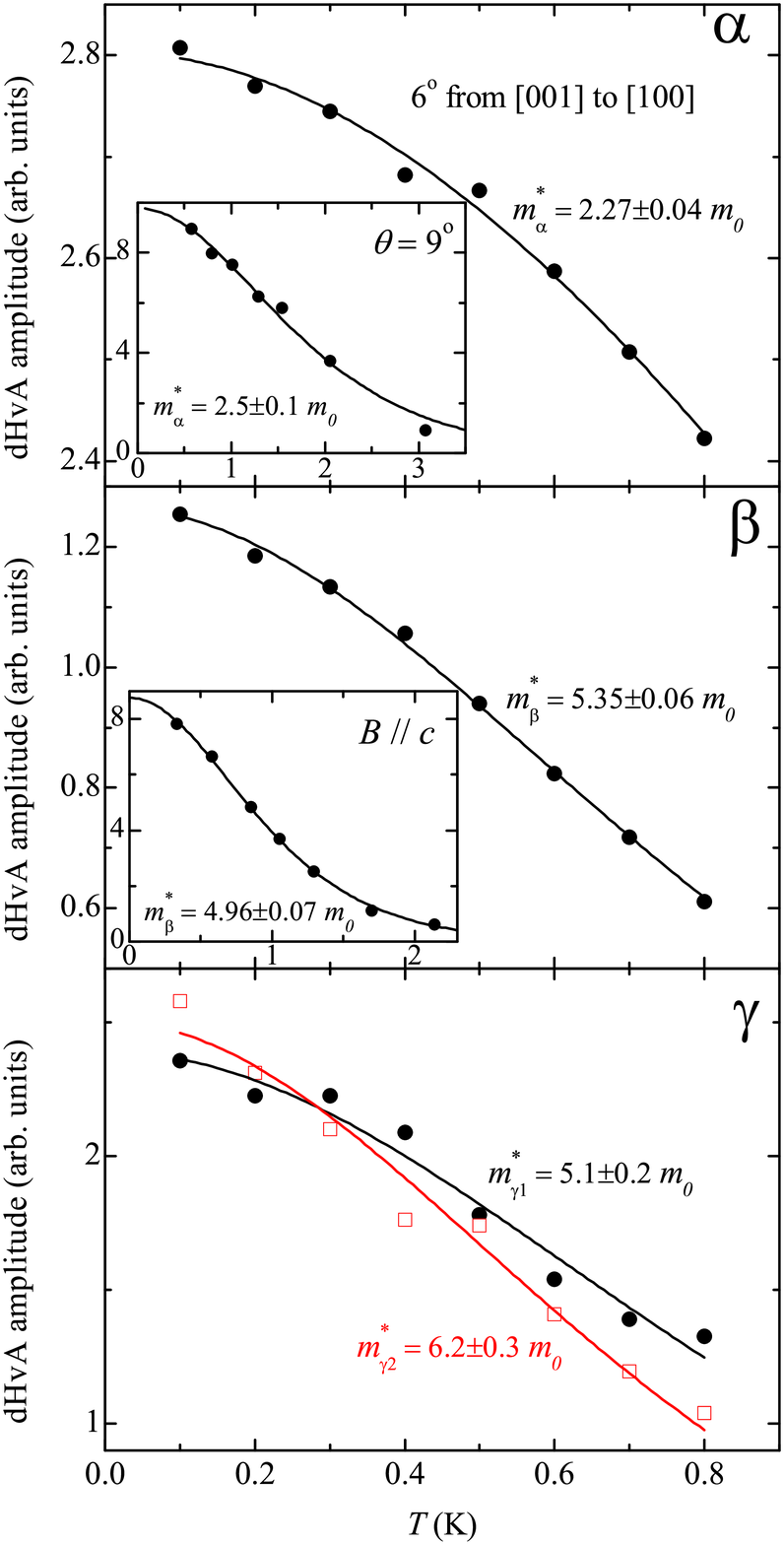}
\caption{\label{effective_mass} Amplitude of the dHvA oscillations in CePt$_2$In$_7$ as a function of temperature for $\alpha$, $\beta$, and $\gamma$ branches that originate from the main quasi-2D FSs. Insets show the data obtained on a smaller sample at different angles and over different temperature ranges. The lines are fits by the standard Lifshitz-Kosevich formula~\cite{Shoenberg2009}.}
\end{figure}

\subsection{Effective mass}

The effective masses of the main FS branches $\alpha$, $\beta$, and $\gamma$ of CePt$_2$In$_7$ were determined by fitting the temperature dependence of the oscillatory amplitude by the standard Lifshitz-Kosevich formula~\cite{Shoenberg2009}, as shown in Fig.~\ref{effective_mass}. This was done for the magnetic field applied at 6$^\circ$ from $c$ to the $a$ axis, in a dilution refrigerator, in the temperature range from 50 to 800 mK and over the field range from 24 to 34.5~T. The results are given in Table~\ref{dHvA parameters}. The effective mass of the $\alpha$ branch is rather small, just slightly over 2$m_0$ ($m_0$ is the bare electron mass). The masses of the $\beta$ and $\gamma$ branches are somewhat higher, ranging from 5$m_0$ to 6$m_0$. As the obtained effective masses are only moderately enhanced, the temperature dependences of the oscillation amplitudes were also measured on a smaller sample in a $^3$He cryostat in the temperature range from 0.33 to about 3 K. These measurements were performed for magnetic field applied parallel to the $c$ axis over the field range from 29 to 33~T and for field applied at 9$^\circ$ from $c$ towards the $a$ axis over the field range from 24 to 33~T. In this case, only a few of the high frequencies had large enough amplitudes to allow the determination of the effective mass. However, the effective masses, which were possible to extract from these higher-temperature measurements, agree very well with the values obtained at a lower temperature and a slightly different angle. The calculated band masses of the thermodynamically important quasi-2D FS sheets of both CePt$_2$In$_7$ and LaPt$_2$In$_7$ for the magnetic field applied at 6$^\circ$ off the $c$ axis are also shown in Table~\ref{dHvA parameters}. The LDA calculations do not take strong electronic correlations into account and provide only the band masses. That is why the calculated masses for CePt$_2$In$_7$ are of the order of a bare electron mass. The effective masses measured in PrPt$_2$In$_7$ are much smaller than the effective masses measured in CePt$_2$In$_7$ but are comparable to the band masses calculated for LaPt$_2$In$_7$. This confirms the expected absence of strong electronic correlations in PrPt$_2$In$_7$.

The effective masses measured in CePt$_2$In$_7$ account reasonably well for this material's moderate specific-heat coefficient $\gamma = 180 \; \rm{mJ/K^2mol}$~\cite{Krupko2016}. This value of $\gamma$ is considerably smaller than the value of $\gamma \approx 400 \; \rm{mJ/K^2mol}$ reported for CeRhIn$_5$~\cite{Cornelius2000,Cornelius2001}, despite the similarities in the effective masses of the two compounds (masses of 3.5$m_0$--6$m_0$ were reported for quasi-2D FSs of CeRhIn$_5$~\cite{Hall2001a,Shishido2002}). However, some of the theoretically predicted FS orbits were not observed experimentally in CeRhIn$_5$, whereas we have observed all the large, dominant FSs in CePt$_2$In$_7$ and PrPt$_2$In$_7$.

Finally, the effective masses we detected in magnetic fields below 35~T in CePt$_2$In$_7$ are somewhat higher than the values reported at much higher magnetic fields, above $B_m \approx$ 45 T~\cite{Altarawneh2011}. This difference is, however, too small to explain why the high dHvA frequencies were not observed below $B_m$ in the previous measurements. It is not clear at present whether the effective masses decrease continuously with magnetic field or change abruptly at $B_m$, whose origin is as yet unknown. Careful low-temperature measurements in pulsed magnetic fields are required to clarify this issue.

\subsection{CePt$_2$In$_7$ Fermi surface dimensionality}

\begin{figure}[htb]
\includegraphics[width=7.5cm]{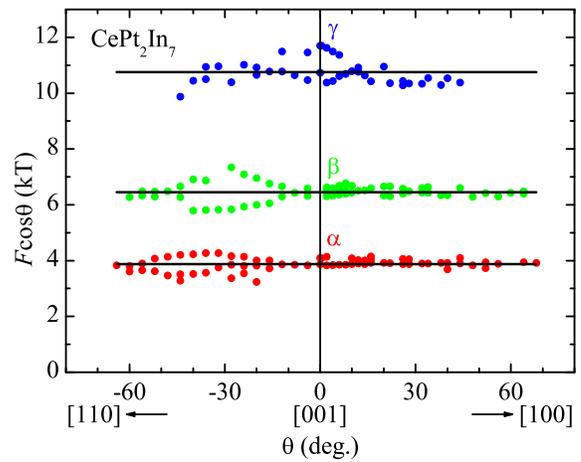}
\caption{\label{AngDep2D} Angle dependence of the experimental dHvA frequencies plotted as $F\cos(\theta)$ for the three main FS sheets of CePt$_2$In$_7$. Solid lines represent the data average of $F\cos(\theta)$ for each FS.}
\end{figure}

As shown in Fig.~\ref{FS}, the major FS sheets calculated for LaPt$_2$In$_7$ are quasi-2D slightly corrugated cylinders extending along the $c$ axis. Given the excellent agreement between the angle dependence of dHvA frequencies measured in CePt$_2$In$_7$ and the results of calculations for LaPt$_2$In$_7$, we can say that the FSs calculated for the latter compound represent the real FSs of CePt$_2$In$_7$. As many of the physical properties of HF materials strongly depend on the FS dimensionality, it is important to estimate to what extent the FSs of CePt$_2$In$_7$ can be approximated by cylinders typical for an ideal 2D system. For an ideal cylindrical FS, the angle dependence of its cross section and, therefore, of the dHvA frequency is given by $F(\theta) = F_0/\cos(\theta)$, where $F_0$ is the dHvA frequency for field applied along the $c$ axis and $\theta$ is the angle between the $c$ axis and the magnetic field. Therefore, $F\cos(\theta)$ should be independent of angle. Figure~\ref{AngDep2D} shows the measured values of $F\cos(\theta)$ plotted as a function of $\theta$ for the three main FSs of CePt$_2$In$_7$. For all three FSs, $F\cos(\theta)$ is almost independent of angle for the magnetic field applied in the [001]-[100] plane, in agreement with previously reported results~\cite{Altarawneh2011}. For the field applied in the [001]-[110] plane, the dispersion is slightly larger, in agreement with calculated FSs shown in Fig.~\ref{FS}. However, the deviation from the average values remains relatively small, below 10\%, thus indicating that the main FS sheets of CePt$_2$In$_7$ are close to being ideal cylinders.

\section{Conclusions}

In summary, we performed dHvA effect measurements in CePt$_2$In$_7$ and PrPt$_2$In$_7$ in high magnetic fields. The observed dHvA frequencies and their angle dependence were found to be almost identical in the two compounds. This implies that the main FSs are nearly the same in these two materials. These main FSs are close to cylindrical, i.e., almost ideally 2D, as is expected from the characteristic layered crystal structure. However, in CePt$_2$In$_7$ the dHvA frequencies corresponding to the large quasi-2D FSs are detected only above a certain angle-dependent threshold magnetic field, which is, however, much lower than $B_m \approx$ 45 T previously reported~\cite{Altarawneh2011}. This calls into question the degree of similarity between CePt$_2$In$_7$ and CeIn$_3$ suggested in the previous work~\cite{Altarawneh2011}. The existence of a threshold field is likely due to the modification of the paramagnetic Brillouin zone by antiferromagnetic ordering. In this scenario, the original paramagnetic FSs can be observed as a result of magnetic breakdown that begins to occur at high magnetic fields.

The comparison of the experimental results with band structure calculations suggests unambiguously that the $f$ electrons are localized and do not contribute to the Fermi volume in CePt$_2$In$_7$. The same conclusion was drawn for both CeIn$_3$~\cite{Ebihara1993,Endo2005} and CeRhIn$_5$~\cite{Shishido2002} deep inside the antiferromagnetic state. In these compounds, however, a FS reconstruction occurs when antiferromagnetism is suppressed by either pressure~\cite{Settai2005,Shishido2005} or magnetic field~\cite{Harrison2007,Jiao2015}. From this point of view, it would be interesting to investigate if and how the FS of CePt$_2$In$_7$ changes across the quantum critical points induced by pressure or magnetic field.

\begin{acknowledgments}
We are grateful to Y. Haga for the determination and sharing with us the details of the crystal structure of PrPt$_2$In$_7$. We acknowledge the support of the HFML-RU/FOM, the LNCMI-CNRS, and the HLD-HZDR, members of the European Magnetic Field Laboratory (EMFL), ANR-DFG grant ``Fermi-NESt,'' and JSPS KAKENHI Grants No. JP15H05882, No. JP15H05884, No. JP15H05886, No. JP15K21732 (J-Physics). K.G. acknowledges support from the DFG within GRK 1621. R.D.H.H. acknowledges financial support from the Radboud Honours Academy of the Faculty of Science.
\end{acknowledgments}

\bibliography{CePt2In7_dHvA}

\end{document}